\documentclass[aps,pra,twocolumn]{revtex4}
\usepackage[usenames]{color}
\newcommand{\comment}[1]{}
\usepackage{graphicx}
\usepackage{amsmath}
\usepackage{epsfig}
\usepackage{eufrak}
\newcommand{\bra}[1]{\left\langle {#1} \right|}
\newcommand{\ket}[1]{\left| {#1} \right\rangle}
\newcommand{\expect}[1]{\langle {#1} \rangle}
\newcommand{\ketn}[1]{\left. \! \! {#1} \right\rangle}
\def\bo{{\bf \Omega}}

\def\Pn{{\cal P}}
\def\e{\mathfrak{e}}
\def\b{\mathfrak{b}}
\def\f{\mathfrak{f}}
\def\be{\begin{equation}}
\def\ee{\end{equation}}

\def\summ{\sum\limits}
\def\ba{\begin{array}}
\def\ea{\end{array}}

\def\prodd{\prod\limits}

\newcommand\bcoh[1]{\bra{( {#1} )^{2F}}}
\newcommand\kcoh[1]{\ket {( {#1} )^{2F}}}
\newcommand{\ketsub}[2]{| {#1}\hspace{1mm} {\rangle}_{{\hspace{-0.5mm}}_{#2}} }
\newcommand{\brasub}[2]{\hspace{1mm}  {\langle}_{{\hspace{-2mm}}_{#2}}\hspace{1mm} {#1} |}
\newcommand{\V}{V_{\rm int}}

\newcommand\co[1]{\left(\Omega_{#1}\right)^{2F}}
\newcommand\cotr[1]{\left(\Omega_{#1}^t\right)^{2F}}

\begin{document}

\title{A geometrical approach to the dynamics of spinor condensates I:~Hydrodynamics}

\date{\today}

\author{Ryan Barnett,$^1$ Daniel Podolsky,$^{2,3}$ and Gil Refael$^{1}$}
\affiliation{$^{1}$Department of Physics, California Institute of
Technology, MC 114-36, Pasadena, California 91125, USA}
\affiliation{$^2$Department of Physics, University of Toronto,
Toronto, Ontario M5S 1A7, Canada}
\affiliation{$^3$Department of Physics, Technion, Haifa, 32000, Israel}

\begin{abstract}
In this work we derive the equations of motion governing the
hydrodynamics of spin-$F$ spinor condensates. 
We pursue a description based on standard physical variables (total
density and superfluid velocity), alongside $2F$ `spin-nodes': unit
vectors that describe the spin $F$ state, and also exhibit the
point-group symmetry of a spinor condensate's mean-field ground state.
The hydrodynamic equations of motion consist of a mass continuity equation, $2F$ Landau-Lifshitz
equations for the spin-nodes, and a modified Euler equation.
In particular, we provide a generalization of the Mermin-Ho relation
to spin one, and find an analytic solution for the skyrmion texture in the
incompressible regime of a spin-half condensate.  These results exhibit a beautiful
geometrical structure that underlies the dynamics of spinor condensates.

\end{abstract}

\maketitle

\section{Introduction}

A central theme of contemporary atomic physics experiment 
is the dynamics of Bose-Einstein
condensates and other correlated atomic gases. Of particular interest
are mixtures of several species such as Fermi-Bose mixtures
\cite{schreck01,hadzibabic02,roati02}, and bosonic gases with an internal
spin
degrees of freedom, {\em i.e.}, spinor condensates.
Spin-one and spin-two spinor condensates have been realized
as particular hyperfine states of alkali
atoms \cite{stenger98,schmaljohann04, chang04}.  In addition
the trapping and cooling of
$^{52}$Cr atoms has led to the realization of
a spin-three condensate \cite{griesmaier05}.
On the theoretical front,
since the initial work of Ohmi and Machida \cite{ohmi98}
and Ho \cite{ho98}
numerous interesting works have followed that
discuss ground states, dynamics,
and topological excitations of such systems (see, for instance,
a review in [\onlinecite{lewenstein06}]).

Recently, it was shown within mean-field theory that the ground states
of spinor condensates exhibit a high degree of symmetry.  This symmetry is
opaque in the standard spinor description of the
condensate. On the other hand, the symmetry is transparent in the so-called reciprocal 
state representation. Here,
one uses the fact that the mean field ground state of a spin-$F$ condensate can be 
described by $2F$ coherent spin states orthogonal to it.
Each one of these so-called  reciprocal states is fully spin-polarized, 
pointing along some
direction on the unit sphere.  Since there are $2F$ such
reciprocal states, the ground state is uniquely described
(up to an overall phase) in terms of the $2F$ points on the unit sphere
\cite{barnett06}.  For typical spinor condensate Hamiltonians, these points,
which we denote ``spin nodes'', form highly symmetric configurations.  
For instance, an $F=2$
condensate has a cyclic phase, where the spin nodes are arranged
in a tetrahedron, as well as a square phase.

The  spin-node description of the ground states
of spinor condensates provides an intuitive
geometrical description of the state of the condensate.  In addition,
it provides a parametrization which readily exhibits the
hidden point-group symmetries of the state. Despite its appeal, however, this
parametrization has not been used to describe the dynamics of
spinor condensates. Our goal in this paper is to provide a complete description of the
hydrodynamics of spinor condensates in terms of such spin-nodes.

The description of the dynamics we provide in this paper is
hydrodynamic in the sense that
it focuses on the low energy dynamics of the system associated with
locally conserved quantities, or with the slow elastic deformation of
spontaneously broken degrees of freedom.
As with any fluid, such a macroscopic hydrodynamic description
is highly illuminating and useful.
Here we derive such a description using the density,
superfluid velocity,  and the
spin-nodes (the $2F$ vectors on the
unit sphere) as our basic degrees of freedom.  In addition to the Euler equations, which describe
mass, momentum, and energy conservation, we obtain
$2F$ Landau-Lifshitz equations for the dynamics of the
spin-nodes.  Furthermore, our derivation gives
a natural generalization of the Mermin-Ho relation which connects the
vorticity in a ferromagnetic spinor condensate with the Pontryagin
index  of the order parameter.

The paper is organized as follows.  Sec.~\ref{Sec:background}
provides general background on spinor condensates, and reviews
recent progress on  the hydrodynamic description of ferromagnetic condensates \cite{lamacraft08}.
In Sec.~\ref{Sec:spinrep} we
present the spin node representation of spinor-condensate degrees
of freedom, and derive several useful identities within this formalism.
In Sec.~\ref{Sec:Fhalf} we proceed to obtain hydrodynamic equations
for the spin-half condensate, using the spin-node formalism,
and find an analytic solution for a skyrmion configuration.
In Secs.~\ref{Sec:Fone} and \ref{Sec:genF}, we derive the general hydrodynamic equations of motion
for the spin-one condensate, and then for an arbitrary spin-$F$ condensate, which includes
a generalization of the Mermin-Ho relation \cite{mermin76}.

The treatment of the spin degrees of freedom in this paper is exact,
and it accounts for the full geometrical
structure of the hydrodynamics of spinor condensates.  But the
precision of the hydrodynamic description here comes at a price: this formalism becomes increasingly complex as the spin
$F$ grows, and, for large $F$, the analysis of its exact form becomes
impractical.  Nevertheless, the equations derived here even for large
$F$ become quite useful in their linearized form. In an accompanying paper \cite{barnett09},  we show how approximate
methods -- such as linearizing the equations of motion about
mean-field solutions -- elucidate the low energy properties
of spinor condensates with arbitrary spin in a powerful and elegant way.

\section{Background}
\label{Sec:background}

\subsection{Hydrodynamics of spinless BECs}

For a single component BEC, it is natural to expect a simple
hydrodynamic description in terms of density and flow velocity.
We take the time-dependent
Gross-Pitaevskii equation (GPE)
as our starting point:
\be
i\partial_t \psi = -\frac{1}{2} \nabla^2 \psi + g \rho \psi
\ee
where $\psi=\sqrt{\rho}e^{i\theta}$ is the macroscopic wave function, and
$\rho=|\psi|^2$ is the density
(here and after for notational simplicity we will use scaled
units).  This equation can be recast into a
the form of  local momentum and mass conservation laws; with the
superfluid velocity ${\bf v}=\nabla \theta$, one obtains
\cite{pethick02}:
\begin{align}
\partial_t \rho = -\nabla (\rho {\bf v})
\;\; ; \;\;
D_t {\bf v}  = - \nabla \left (g \rho -
\frac{\nabla^2 \sqrt{\rho}}{2\sqrt{\rho}}\right)
\end{align}
where $D_t=\partial_t + {\bf v}\cdot \nabla$ is the material
derivative.  The first of these
is the mass continuity equation, while  the second is
the Euler equation for a fluid, where a quantum pressure
term appears.

\subsection{The hydrodynamics of ferromagnetic BECs and the Mermin-Ho
  relation}

In a series of recent experiments, the quench dynamics of a ferromagnetic
spin-one  condensate was explored 
\cite{sadler06,leslie08,vengalatorre08, vengalatorre09}. 
These
experiments motivated Lamacraft to develop a hydrodynamic
framework for the ferromagnetic BEC in terms of the superfluid
velocity, and the
director of its ferromagnetic order ${\bf n}$ \cite{lamacraft08}.  This
description is particularly illuminating when considering the
instabilities of the system.  This problem was also
theoretically considered in  
Refs.~\cite{saito07,lamacraft07,uhlmann07,damski07,mukerjee07,mias08,cherng08}.

The GP Lagrangian density describing such a ferromagnetic
spinor condensate is given by
\be
{\cal L}= i\psi_a ^{*}\partial_t \psi_a  -\frac{1}{2}
\nabla \psi_a^{*}\cdot\nabla \psi_a - \frac{1}{2} g \rho^2
-\frac{1}{2} c_2 \rho^2  m^2
\label{FMGP}
\ee
where $a=-F,\ldots,F$ is summed over all $F_z$ eigenstates, and
\be
\rho=\sum_a\psi_a^{*}\psi_a \hspace{4mm} {\bf
  m}=\frac{1}{\rho}\sum_{ab}\psi_a^{*} {\bf F}_{ab} \psi_b.
\ee
Lamacraft's approach assumed an incompressible liquid with a
wavefunction restricted to the ferromagnetic phase (assuming large
$g$ and $c_2$)
\be
\psi_a=e^{i\theta}\Phi_a({\bf n})
\ee
where $\Phi_a({\bf n})$ is the highest eigenstate of ${\bf
  n}\cdot {\bf F}$. A substitution of this wavefunction into
Eq. (\ref{FMGP}) yielded the following set
of hydrodynamic equations \cite{lamacraft08}:
\begin{align}
\label{Eq:lam}
\nabla \cdot {\bf v} &=0 \\ \notag
D_t {\bf n} &= \frac{1}{2} {\bf n} \times \nabla^2 {\bf n}.
\end{align}
Once the density is eliminated,
we notice that the spin dynamics are given by a Landau-Lifshitz
equation with the material derivative
$D_t = \partial_t + {\bf v}\cdot \nabla$.
In addition, the vorticity is related to  the
Pontryagin density by
\begin{equation}
\nabla \times {\bf v} = \frac{F}{2} \varepsilon_{\alpha \beta \gamma}
n_\alpha
\left(\nabla n_\beta \times \nabla n_\gamma \right).
\end{equation}
This identity is widely-known as the Mermin-Ho
relation~\cite{mermin76,ho98}.
Among other things, such a relation has important
consequences for the topological defects
in ferromagnetic condensates 
\cite{ho98, mizushima02, mueller04, mizushima04}.
Such hydrodynamic equations were also derived in \cite{stone96}
to describe magnetic properties of quantum hall systems.

Making use of this simple description, Lamacraft showed
that the helical configuration of the ferromagnetic condensate 
\cite{vengalatorre08}
is unstable. In general,
it is clear that such a geometric description simplifies, at least
conceptually, the analysis of spinor-condensate dynamics.

\subsection{General magnetic ground state of
spinor condensates, and the reciprocal
state representation}
\label{sec:reciprocal}

A general spin-$F$ spinor-condensate is described by a macroscopic
wave-function with $2F+1$ complex components $\psi_a$
($a=-F,\ldots,F)$. When quantum fluctuations are unimportant,
the condensate dynamics is described by the
time-dependent Gross-Pitaevskii equation
\begin{equation}
i\partial_t \psi_a = -\frac{1}{2} \nabla^2 \psi_a + 
\frac{\partial V}{\partial \psi^*_a}
\end{equation}
where 
$V$ is the spin-dependent interaction energy.  The
interaction energy is
given by the set of
parameters $g_S$, with $S=0,2,\ldots,2F$ describing
the two-particle interaction
strength in the $S$ total angular momentum channel:
\begin{equation}
\label{Eq:spinint}
\V= \frac{1}{2} \sum_{S,m} g_S \psi_a^* \psi_b^* \bra{ab}\ketn{Sm}
\bra{Sm} \ketn{a'b'}  \psi_{a'} \psi_{b'}.
\end{equation}
In the above, $\bra{ab}\ketn{Sm}$ are Clebsch-Gordan coefficients.
Note that this expression can also be written as the expectation
value of an operator:
\begin{equation}
\label{Eq:spinintalt}
\V=\frac{1}{2}\; \brasub{\psi}{1} \brasub{\psi}{2} {\cal \V}  
\ketsub{\psi}{2} \ketsub{\psi}{1}
\end{equation}
where 
\begin{equation}
{\cal \V} = \sum_{S,m} g_S \ket{Sm} \bra{Sm} = \sum_{S} g_S {\cal P}_S.
\end{equation}
In this expression, ${\cal P}_S$ projects into the total spin $S$ scattering
channel.

The classical (mean-field) ground states occur for uniform condensates
which minimize Eq.~(\ref{Eq:spinint}) for fixed density $\rho$.
This minimization was carried out for $F=1$ \cite{ho98,ohmi98}, $F=2$
\cite{ciobanu00}, and $F=3$ \cite{diener06,santos06} yielding a
multitude of magnetic phases, which minimize $\V$ in different
regions of $\{g_S\}$ parameter space, only one of which (for every
$F$) is ferromagnetic.

Indeed, quite generally, a spin-$F$ spinor condensate may exhibit several
flavors of paramagnetic rather than
ferromagnetic behavior in its ground state. For example, a spin-one
condensate may exhibit the so-called nematic phase,
where $\psi_1=\psi_{-1}=0$, and $\psi_0=1$.
The expectation value of the magnetization for such
a state is clearly zero along any direction,
$\expect{{\bf F}\cdot {\bf n}}=0$. But in the absence of a ferromagnetic
director, ${\bf n}$,
can we still describe a spinor condensate's magnetic state
geometrically?

Such a geometrical method was put forward in
Ref. \cite{barnett06}, based on the use of spin-coherent states.  
A spin-coherent state  $\ket{\Phi_{\bf n}}$ is the eigenvector of 
the operator ${\bf F}\cdot{\bf n}$ with the largest eigenvalue.  
The method of Refs.~\cite{barnett06} relies on 
finding the set of $2F$ spin-coherent states, 
$\{\ket{\Phi_{{\bf n}_i}}\}_{i=1}^{2F}$, which
annihilate the ground state of a uniform condensate:
\be
\bra{\Phi_{{\bf n}_i}}\psi_{GS}\rangle=0.
\label{RS}
\ee
The $2F$ states $\ket{\Phi_{{\bf n}_i}}$ 
provide (up to an overall phase) a unique
description of the magnetic spin-state of the condensate at
each point in space.
Such reciprocal spinors give a natural generalization of the
ferromagnetic director to the case of paramagnetic
condensates. Instead of the geometrically opaque $2F+1$ complex
numbers $\psi_a$, it allows a description of the magnetic state in
terms of $2F$ unit vectors, ${\bf n}_i$, or points on the unit sphere.

\begin{figure}
\includegraphics[width=3.5in]{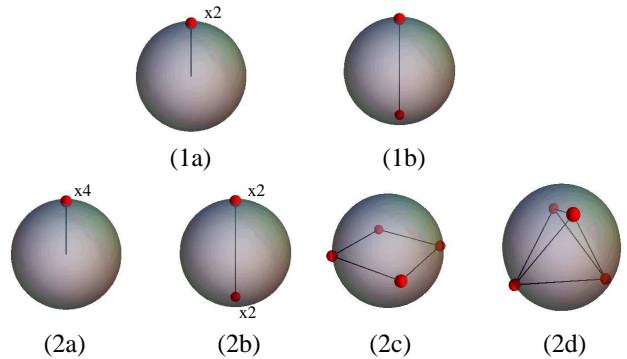}
\caption{Possible phases for the spin-one and spin-two condensates.
Spin-one condensates have ferromagnetic (1a) and nematic phases (1b) while
spin-two condensates have  ferromagnetic (2a), uniaxial nematic
(2b), square biaxial nematic (2c), and a tetrahedral (2d) phases. }
\label{Fig:shapes}
\end{figure}

In addition to its geometrical transparency, such a description also
reveals the highly symmetric nature of the mean-field ground states. All the
paramagnetic phases found so far correspond to a spin node
configuration which is invariant under point
symmetry group operations, and sometimes under a larger symmetry. 
The nematic phase of the $F=1$ condensate, for instance, is described by
two antipodal spin nodes. $F=2$ condensates can exhibit a nematic
phase as well, but also a phase in which the spin-nodes are the
vertices of a square, and a phase with the spin nodes at the vertices of
a tetrahedron.  Such phases are illustrated in Fig.~\ref{Fig:shapes}.

The reciprocal-spinor description was so far only
utilized to discuss equilibrium properties of spinor condensates.
The remarkable geometrical properties and hidden symmetries of
the mean-field ground state, however, provide ample motivation for
employing the spin nodes to obtain a complete
description of the dynamics of spinor condensates. In the following
sections we will develop the tools necessary for such a description,
and use them to derive both a hydrodynamic description, as well as
small oscillation dynamics near  mean-field ground states.

\section{Spin node description of spin-$F$ magnetic states}
\label{Sec:spinrep}

The reciprocal-spinor states so far define the spinor condensate's
state only implicitly through Eq. (\ref{RS}). In order to be able to
use these variables directly, we must invert the relationship, expressing
the spinor condensate Lagrangian directly in terms of these variables.
In order to find this direct representation, we first separate the
wave function into a piece corresponding to the overall density and
phase, and a piece describing the local spin state. We write
\be
\psi_a = \psi \chi_a
\ee
where $\chi_a$ is a normalized spin-$F$ spinor
\be
\sum_a \chi_a^* \chi_a = \bra{\chi}\ketn{\chi}=1
\ee
and  the superfluid density is
\be
\rho=|\psi|^2.
\ee

\subsection{Symmetrized spin-node representation}

As discussed above, a spin-$F$ spinor $\ket{\chi}$ can be described by $2F$ reciprocal states.  On the other hand, such a state can also be described by a fully symmetrized collection of $2F$
spin-half states.  Each spin-half state can be parametrized in terms
of two coordinates $\Omega=(\theta,\phi)$ on the unit sphere,
\be
\ket{\Omega}=\cos \left(\frac{\theta}{2}\right)e^{i\phi/2}\ket{\uparrow}
+\sin\left(\frac{\theta}{2}\right)e^{-i\phi/2}\ket{\downarrow}.
\ee
In this representation,
\be
\ket{\chi}=\frac{1}{\cal{N}}\summ_{\{\sigma\}}\left(\prodd_{i=1}^{2F}
\otimes\ket{\Omega_{\sigma_i}}\right)=\frac{\sqrt{(2F)!}}{\cal{N}}\ket{{\bf \Omega}}
\label{spinsym}
\ee
where $\cal{N}$ is a normalization constant,
and the sum over ${\sigma}$ runs over the $(2F)!$ permutations of the
$2F$ labels for the spin-half parts \cite{majorana45}.  In Eq. (\ref{spinsym}) we also
defined $\ket{\bo}$ as the (unnormalized) sum over
permutations of the tensor product.

The properties of the above formulation are most easily understood using the
Schwinger Boson construction \cite{schwinger65}
(for review, see \cite{auerbach94}). Schwinger
Bosons provide an easy way to construct the Hilbert space of a spin-$F$
spinor state.
We define two Schwinger boson creation operators:
$\hat{a}^\dagger,\hat{b}^\dagger$. An $\hat{a}^\dagger$ boson adds $1/2$ to both the total
spin, and to $F_z$, whereas a $\hat{b}^\dagger$ boson adds $1/2$ to the total
spin, but lowers $F_z$ by half. 
In this notation
\begin{align}
F_{\rm tot}&=\frac{1}{2}(\hat{a}^\dagger \hat{a} + \hat{b}^\dagger \hat{b});\notag\\
F_x&=\frac{1}{2}(\hat{a}^\dagger \hat{b} + \hat{b}^\dagger \hat{a});
\notag
\\
F_y&=\frac{1}{2i}(\hat{a}^\dagger \hat{b} - \hat{b}^\dagger \hat{a});
\label{Eq:schwingerbosonspins}
\\
F_z &= \frac{1}{2} (\hat{a}^\dagger \hat{a} - \hat{b}^\dagger
\hat{b}). \notag
\end{align}
A spin-half spinor is written as:
\be
\ket{\Omega}=u\ket{\uparrow}+v\ket{\downarrow}\nonumber
=(u \hat{a}^\dagger+v \hat{b}^\dagger)\ket{0}\nonumber
\ee
with $\ket{0}$ the Schwinger-boson vacuum.  Here, $u$ and $v$ can be written in terms of the
coordinates on the unit sphere as $u=\cos(\theta/2)e^{i\phi/2}$ and $v=\sin(\theta/2)e^{-i\phi/2}$.

A symmetrized tensor product of $2F$ spins within the SB formalism is
simply written as:
\be
\ket{\bf \Omega}=\ket{\Omega_1 \ldots \Omega_{2F}}
= \prod_{i = 1}^{2F} \left(u_i \hat{a}^{\dagger} +
v_i \hat{b}^{\dagger}\right) \ket{0}.
\label{Eq:spinor}
\ee
with $u_i$ and $v_i$ parametrized in terms of $\theta_i,\,\phi_i$ 
as shown above.  We refer to this collection of the $2F$ spin-half
states which construct $\ket{\bo}$ as ``spin nodes''.

If we wish to calculate wavefunction overlaps using the Schwinger
Boson formalism, we can use Wick's theorem to obtain
\be
\ba{l}
\langle{\bo}^{(a)}\ket{\bo^{(b)}}=\\
=\bra{0}\prod_{i = 1}^{2F} \left(
u_i^{(a)*}\hat{a} +v_i^{(a)*}\hat{b}\right)\prod_{j = 1}^{2F} \left(
u_j^{(b)}\hat{a}^{\dagger} +v_j^{(b)}\hat{b}^{\dagger}\right)\ket{0}
\\
=\summ_{\{\sigma\}}\prodd_{i=1}^{2F}\bra{0} \left(
u_i^{(a)*}\hat{a} +v_i^{(a)*}\hat{b}\right) \left(
u_{\sigma_i}^{(b)}\hat{a}^{\dagger} +
v_{\sigma_i}^{(b)}\hat{b}^{\dagger}\right)\ket{0}\\
=
\summ_{\{\sigma\}}\prodd_{i=1}^{2F}\langle\Omega_i^{(a)}\ket{\Omega_{\sigma_i}^{(b)}}.
\ea
\ee
where $\sigma$ is a permutation of the $2F$ indices that mark the
spin-half parts. This result could have also been obtained directly
from Eq. (\ref{spinsym}). Nevertheless, we find it instructive to
demonstrate the simple Schwinger Boson construction to obtain the
symmetrized states.

\subsection{Connection between spin nodes and reciprocal spinors}

Since the symmetrized  spin-node representation
can be used to express any
spin state directly, it makes a grossly overcomplete
basis. Nevertheless, its usefulness arises since it perfectly reflects
the spin-nodes formalism of the spinor-condensates ground
states \cite{barnett06,barnett07}. In the following we will introduce
the necessary new notation for the spin-node formalism; we summarize
the new notation in Appendix \ref{A1}

It is simple to see that a spin-coherent state can be written in terms of Schwinger boson
states as
\begin{eqnarray}
\kcoh{\Omega} &=& \left(u \hat{a}^{\dagger} +v
\hat{b}^{\dagger}\right)^{2F}\ket{0}\label{Eq:spincoherent}\\
&=&\ket{\Omega\ldots\Omega}\nonumber.
\end{eqnarray}
Thus a coherent state can be thought of as $2F$ spin-nodes
pointing in the same direction.  (For a summary of the notation see Appendix \ref{A1}.)

As described in Sec.~\ref{sec:reciprocal}, a reciprocal spinor is a coherent state
$\kcoh{\Omega_r}=\ket{\Omega_r \ldots \Omega_r}$ orthogonal to a given
spinor $\ket{\bo}$.
Using the construction in terms of symmetrized spin nodes, we can write an
equation to determine the reciprocal spinors for a particular state
$\ket{\bo}=\ket{\Omega_1 \ldots \Omega_{2F}}$:
\be
\label{eq:reciprocal}
\bcoh{\Omega_r} \ketn{\bo}
=(2F)! \prod_{i=1}^{2F} \bra{\Omega_r}\ketn{\Omega_i}=0.
\ee
This equation has $2F$ solutions, each corresponding to a different
term in the product vanishing. That is, the $ith$ solution of
Eq. (\ref{eq:reciprocal}) is
\be
\kcoh{\Omega_r}=\kcoh{\Omega_i^t} = \ket{\Omega_i^t \ldots \Omega_i^t},
\ee
where
\be
\ket{\Omega_i^t}
=(-v_i^*\hat{a}^{\dagger}+u_i^*\hat{b}^{\dagger})\ket{0}
\ee
is the time-reversed
spinor of $\ket{\Omega_i}$   with
$\theta_i^t=\pi-\theta_i,\,\phi_i^t=\phi_i+\pi$.  Here we use the fact that a spin-half spinor is orthogonal to its time-reversed counterpart.

The set of reciprocal coherent states is, therefore, just the set of
states antipodal to the directions in the spin-node set $\{\Omega_i\}_{i=1}^{2F}$ on the unit
sphere. Using this construction,
Eq. (\ref{Eq:spinor}) provides us with a direct
expression for a spinor condensate state in terms of its reciprocal
vectors.

\subsection{Time derivatives of spinors and spin-nodes}

As stated above, our goal in this paper is to extract the dynamics
in terms of individual spin nodes $\ket{\Omega_i}$.  In order to do so,
we must be able to isolate the dynamics of each spin node within $\ket{\bo}$.
Consider the time derivative of $\ket{\bo}$ which will appear in the
GPE. We can express it as a sum of terms in which the time derivative
operates on individual spin nodes:
\begin{equation}
\partial_t \ket{\bo} = \ket{\partial_t \Omega_1 \Omega_2 \ldots}
+ \ket{\Omega_1 \partial_t \Omega_2 \ldots} + \ldots.
\end{equation}
The trick that allows us to isolate individual spin nodes consists 
of taking the inner product
of $\partial_t\ket{\bo}$ with the $ith$ reciprocal-state of $\ket{\bo}$,
which is $\kcoh{\Omega_i^t}$.
All terms which do not involve a time derivative of $\ket{\Omega_i}$
identically vanish and we are left with the single term
\begin{equation}
\bcoh{\Omega_i^t} \partial_t \ket{\bo}= (2F)!\bra{\Omega_i^t}
\ketn{\partial_t \Omega_i} \prod_{\ba{c}j=1\\j\neq i\ea}^{2F} \bra{\Omega_i^t} \ketn{\Omega_j}.
\label{dttrick}
\end{equation}
We will make extensive use of this method
for isolating the dynamics of individual spin nodes in the following sections.

\subsection{Geometrical parametrization of the spin-half components: Moving
from $\ket{\Omega_i}$ to ${\bf n}_i$}\label{Sec:vecs0}

All results above were concerned with breaking a spin-$F$ spinor into
its $2F$ spin-half parts, $\ket{\Omega_i}$, and with the correspondence
between these spin-half parts and the reciprocal coherent states.
We would like, however, to understand
the dynamics of spinor condensates not only in terms of the spinors,
$\ket{\Omega_i}$, but also in terms of the unit vectors that describe
them, ${\bf n}_i$, where $\ket{\Omega_i}$ is highest value eigenvector
of ${\bf F}\cdot {\bf n}_i$.

The first step in finding the equations of motion in terms of the spin-directors
${\bf n}_i$, is to establish an orthonormal triad
$({\bf e}_x, {\bf e}_y, {\bf n})$
that parameterizes  
the space on $S^2$ in the vicinity of ${\bf n}_i$. In the following we
will only consider a single spin-half part, and therefore we drop the
index $i$.

The first element of the triad is  ${\bf n}$ itself:
\be
{\bf n} = 2 \bra{\Omega}
{\bf F} \ket{\Omega}
\ee
where ${\bf F}$ is the spin operator, acting in the spin-half Hilbert
space. To complete the triad,
we again use the time-reversed spin-half
ket, $\ket{\Omega^t}$ where $\bra{\Omega^t}\ketn{\Omega}=0$.
With this, we can construct states pointing
in the ``$x$'' and ``$y$'' directions with respect to ${\bf n}$ as
\begin{align}
\ket{\Omega_x}&= \frac{1}{\sqrt{2}}(
\ket{\Omega}+\ket{\Omega^t})\\
\ket{\Omega_y}&=
\frac{1}{\sqrt{2}}( \ket{\Omega}+i \ket{\Omega^t}).
\end{align}
These states allow us to complete the orthonormal triad by
defining
\be
{\bf e}_x = 2 \bra{\Omega_x} {\bf F} \ket{\Omega_x},\hspace{1cm}
{\bf e}_y = 2 \bra{\Omega_y} {\bf F} \ket{\Omega_y}.
\ee
From these we can construct
\begin{equation}
{\bf e}_{\pm} = {\bf e}_x \pm i{\bf e}_y.
\end{equation}
It is useful to note that ${\bf F} \cdot {\bf e}_{\pm}$ act
as raising and lowering operators.  That is,
\begin{equation}
{\bf F} \cdot {\bf e}_+ \ket{\Omega}=0 \; ; \;
{\bf F} \cdot {\bf e}_+ \ket{\Omega^t}=\ket{\Omega},
\end{equation}
with similar relations holding for lowering operators.

Note that there is an
ambiguity in such coordinate systems since ${\bf e}_{x}$ and
${\bf e}_{y}$ can together be rotated about ${\bf n}$ which
corresponds to the gauge choice for the spinors.  That is,
the gauge of a spinor can be changed by
$\ket{\Omega} \rightarrow e^{i\lambda}\ket{\Omega}$ without changing
its direction ${\bf n}$.
In general, quantities
which are gauge invariant cannot depend on the parameterization of
the spin and will only involve the unit vectors ${\bf n}_i$.
We will adhere to the following gauge convention
for parameterizing  spin-half spinors:
\be
\ket{\Omega}=\cos(\theta/2)e^{i\phi/2}\ket{\uparrow}+\sin(\theta/2)e^{-i\phi/2}\ket{\downarrow}
\ee
In this gauge-choice it is easy to see that:
\be
{\bf e}_x=\hat{\bf \theta} \hspace{0.5cm} {\bf e}_y=\hat{\bf \varphi}
\ee
where $\hat{\bf \theta}$ and $\hat{\bf\varphi}$ are unit vectors from the
spherical coordinate system.

To complete the discussion, we
make two observations that will simplify the following analysis.
First, we express
${\bf F}$ in the basis of our triad as
\begin{align}
\notag
{\bf F} &= ({\bf F} \cdot {\bf n}) {\bf n}+ ({\bf F} \cdot {\bf
e}_x) {\bf e}_x + ({\bf F} \cdot {\bf e}_y) {\bf e}_y \\
&= ({\bf F}
\cdot {\bf n}) {\bf n}+ \frac{1}{2}({\bf F} \cdot {\bf e}_+) {\bf
e}_- + \frac{1}{2}({\bf F} \cdot {\bf e}_-) {\bf e}_+
\label{Eq:Fbasis}
\end{align}
In addition we note that we can use the spin operator ${\bf F}$
as a projection onto $\ket{\Omega}$ and its time-reversed partner
$\ket{\Omega^t}$ by
\begin{equation}
\ket{\Omega} \bra{\Omega} =
\frac{1}{2} + {\bf n} \cdot {\bf F}
\label{id1}
\end{equation} 
and
\begin{equation}
\ket{\Omega^t}
\bra{\Omega^t} = \frac{1}{2} - {\bf n} \cdot {\bf F}.
\label{id2}
\end{equation}

\subsection{Derivatives of spin-half spinors in terms of the triad
$(\,{\bf e}_x,\,{\bf e}_y,{\bf n})$}

\label{Sec:derivatives}

The relations derived and recalled in the previous sections allow us
to also write derivatives of spinors in terms of vector quantities and
their derivatives.
The terms that we will encounter arise from terms such as the
isolated time derivatives in Eq. (\ref{dttrick}).
Let us now find this decomposition in terms of the
triad $({\bf e}_x,\,{\bf e}_y,{\bf n )}$ and its differential
forms.

Our goal is thus to find:
\be
a_{\alpha}=i\bra{\Omega}\ketn{\partial_{\alpha}\Omega} \;\;{\rm and}\;\;
 \bra{\Omega^t}\ketn{\partial_{\alpha}\Omega}
\label{az}
\ee
in terms of $({\bf e}_x,\,{\bf e}_y,{\bf n})$ and their derivatives. 
We define $a_{\alpha}$ in this form
for reasons that will become clear later.
The first object in Eq. (\ref{az}) can be found by considering the quantity
\be
\partial_{\alpha}(\bra{\Omega^t}{\bf e}_-\cdot
{\bf F}\ket{\Omega})=\partial_{\alpha} 1 =0.
\ee
Allowing the derivative to operate on the bra, the ket, and the vector
${\bf e}_-$, we find
\be
\bra{\partial_{\alpha}\Omega^t}\ketn{\Omega^t}+\bra{\Omega}\ketn{\partial_{\alpha}\Omega}=-\frac{1}{2}{\bf
  e}_+\cdot\partial_{\alpha}{\bf e}_-
\ee
On the left-hand side we used the facts that
${\bf F} \cdot {\bf e}_{-} \ket{\Omega}=\ket{\Omega^t}$ and
$\bra{\Omega^t}{\bf F} \cdot {\bf e}_{-}=\bra{\Omega}$.
On the right-hand side we used the fact that
$\bra{\Omega^t} {\bf F} \ket{\Omega}=\frac{1}{2} {\bf e}_+$ which
can be verified from Eq.~(\ref{Eq:Fbasis}).
It is easy to
verify that
\[
\ba{c}
\bra{\partial_{\alpha}\Omega^t}\ketn{\Omega^t}=\bra{\Omega}\ketn{\partial_{\alpha}\Omega},\\
{\bf e}_x\cdot \partial_{\alpha}{\bf e}_y=-{\bf e}_y\cdot \partial_{\alpha}{\bf e}_x
\ea
\]
from which we find
\be
a_{\alpha}=\frac{1}{2} {\bf e}_y \cdot \partial_\alpha {\bf e}_x
\ee
which is the desired result.

To obtain $\bra{\Omega^t}\ketn{\partial_\alpha \Omega}$ 
we use a similar trick.  Starting with
\be
0=\bra{\Omega^t}\ketn{\Omega^t}\bra{\Omega^t}\ketn{\Omega}
=\bra{\Omega^t}\frac{1}{2}
- {\bf n} \cdot {\bf F}\ket{\Omega}
\ee
we find
\be
\partial_{\alpha}(\bra{\Omega^t}\frac{1}{2}
- {\bf n} \cdot {\bf F}\ket{\Omega})=0.
\ee
As before, allowing the differentiation to act on the bra, the ket,
and ${\bf n}$ results in
\be
\bra{\Omega^t}\ketn{\partial_{\alpha}\Omega}=\bra{\Omega^t}(\partial_{\alpha}{\bf n}) \cdot {\bf F}\ket{\Omega}
\ee
where the term with $\bra{\partial_{\alpha}\Omega^t}$ vanishes since
$(\frac{1}{2}- {\bf n} \cdot {\bf F})\ket{\Omega}=0$. Now, using again the
decomposition in Eq. (\ref{Eq:Fbasis}), we readily find
\begin{equation}
\bra{\Omega^t}\ketn{\partial_\alpha \Omega}= \frac{1}{2}
{\bf e}_+ \cdot \partial_\alpha {\bf n}.
\end{equation}

This concludes all the tools we will need for our analysis below. We
have found how to directly write a spin-$F$ spinor in terms of its spin-nodes,
and extract terms having to do with individual spin nodes
out of sums arising, e.g., from differentiation.  
Furthermore we translated the spin-half
representation of $\ket{\Omega_i}$ to a set of $2F$ triad bases 
$({\bf e}_{ix}, {\bf e}_{iy}, {\bf n}_i)$ which will allow us to 
parametrize the spin state geometrically. Appendix \ref{Notation}
summarizes the various notation introduced throughout this section.

\section{Hydrodynamics of Spin-half condensates}
\label{Sec:Fhalf}

One of our main goals is  to write the
exact (mean-field) equations of motion for a spinor condensate in terms
of the spin nodes and the superfluid velocity and density. In this
section we achieve this goal for spin-half condensates. 
The equations of motion can be trivially generalized to general 
spin-$F$ condensates restricted to the ferromagnetic state, when the
spinor $\ket{\chi}$ is restricted to be a coherent spin-state. In this
case the equations of motion for the condensate reduce to those we
find below.

\subsection{Gross-Pitaevskii Lagrangian}

In this section we consider the Gross-Pitaevskii 
Lagrangian. We begin by writing the Lagrangian in a revealing
form, using the representation of the bosonic field which separates
the spinor order parameter into a product of a density piece and a spin
piece, $\psi_a = \psi \chi_a$. The GP Lagrangian is then:
\begin{align}
{\cal L}&= \notag
i \psi_a^{*} \partial_t \psi_a - \frac{1}{2}
\nabla \psi_a^*\cdot \nabla \psi_a
- \V. \\
&=i \psi ^{*} \partial_t \psi +  \rho a_t
- \frac{1}{2} |(-i \nabla -{\bf a})\psi|^2 - \frac{1}{2} \rho
\Upsilon- \V.
\label{GP1}
\end{align}
where $\V$ is the spin-related interaction and $\rho=|\psi|^2$
with $\psi$ a complex field.  Eq. (\ref{GP1}) defines the spin vector
potential:
\begin{equation}
a_t \equiv i\bra{\chi}\ketn{\partial_t\chi} \; ; \;
{\bf a} \equiv i \bra{\chi}\ketn{\nabla \chi},
\label{VP}
\end{equation}
and the quantity
\begin{equation}
\Upsilon \equiv \bra{\partial_\alpha \chi} \ketn{\partial_\alpha \chi} - \bra{\chi} \ketn{\partial_\alpha \chi}
\bra{\partial_\alpha \chi} \ketn{\chi}.
\label{Yps}
\end{equation}
An
interesting observation is that the quantity $\Upsilon$ for a
general spin $F=N/2$ can be identified with the CP$^N$ model
from quantum field theory \cite{rajaraman82}.
Notice that there is a $U(1)$ gauge freedom in the density-spin
decomposition:
\[
\psi\rightarrow e^{i\lambda}\psi ,\, \ket{\chi}\rightarrow
e^{-i\lambda}\ket{\chi}.
\]
The quantity $\Upsilon$, however, is gauge 
independent. We make a gauge choice when we write
the normalized $\ket{\chi}$ as in Eq. (\ref{spinsym}), with $\ket{\bo}$
written as Eq. (\ref{Eq:spinor}). This $U(1)$ gauge freedom is also
reflected in an ambiguity in the choice of the triad arising from the spin-half parts
of $\ket{\chi}\propto\ket{\bo}$, since for each spin-part ${\bf e}_{x}$ and
${\bf e}_{y}$ can together be rotated about ${\bf n}$. The choice of a
particular triad is set by the gauge choice. In general, quantities
which are gauge invariant cannot depend on the parametrization of
the spin, and will only involve the unit vectors ${\bf n}_i$.
The vector potential can be related to the superfluid velocity
by
\begin{equation}
{\bf v}= \frac{1}{\rho 2i}(\psi_a^* \nabla \psi_a - \psi_a \nabla \psi_a^*)
= \nabla \theta - {\bf a}
\end{equation}
where $\theta$ is the argument of $\psi$.
So far we have not used the fact that the spin is $F=1/2$.

\subsection{Geometric representation of hydrodynamic quantities}
\label{Sec:vecs}

Now that we know the quantities of interest in the spinor description
of the GP Lagrangian, we can translate them to the hydrodynamic
variables of density and magnetization direction.

The most important quantity appearing above is the vector potential as
defined in Eq.~(\ref{VP}). Following the discussion in
Sec. \ref{Sec:vecs0} we see that for a spin-half condensate the vector potential is
\begin{equation}
a_{\alpha}= i \bra{\Omega}\ketn{ \partial_\alpha \Omega} =
\frac{1}{2} {\bf e}_y \cdot \partial_\alpha {\bf e}_x.
\end{equation}
The analogy between ${\bf a}$ and the vector potential appearing in
the Maxwell equations compels us to
consider the antisymmetric field tensor
$\f_{\alpha \beta}= \partial_\alpha a_\beta - \partial_\beta a_\alpha$.
Through a series of manipulations this can be written purely in
terms of $\bf n$
\begin{eqnarray}
\f_{\alpha \beta} &=& \frac{1}{2} \partial_\alpha {\bf e}_y \cdot
\partial_\beta {\bf e}_x-\frac{1}{2} \partial_\beta {\bf e}_y \cdot
\partial_\alpha {\bf e}_x
\\
&=& \nonumber
\frac{1}{2}( \partial_\alpha {\bf e}_y \cdot {\bf n})
(\partial_\beta {\bf e}_x \cdot {\bf n}) -
\frac{1}{2} (\partial_\beta {\bf e}_y \cdot {\bf n})
(\partial_\alpha {\bf e}_x \cdot {\bf n} )
\\
&=& \nonumber
 \frac{1}{2}  ({\bf e}_y \cdot \partial_\alpha{\bf n})
({\bf e}_x \cdot \partial_\beta {\bf n}) -
\frac{1}{2}({\bf e}_y \cdot  \partial_\beta  {\bf n})
({\bf e}_x \cdot \partial_\alpha  {\bf n})\\
&=& \nonumber
\frac{1}{2} ({\bf e}_y \times {\bf e}_x) \cdot (  \partial_\alpha{\bf n}
\times \partial_\beta{\bf n})
\\
&=& \nonumber
-\frac{1}{2} {\bf n} \cdot (  \partial_\alpha{\bf n}
\times \partial_\beta{\bf n}).
\end{eqnarray}
Note that in the above we have repeatedly used the fact that
${\bf v}\cdot \partial {\bf v} =0$ for any unit vector ${\bf v}$.
The result is the Pontryagin topological density, 
which is the object of the celebrated Mermin-Ho relation for spin-half spinors
\cite{mermin76,kamien02}.

The only remaining term is the gauge invariant quantity
$\Upsilon$, defined in Eq. (\ref{Yps}). For a spin-half state, we find:
\begin{align}
\Upsilon &=
\bra{\partial_\alpha \Omega}\ketn{\partial_\alpha \Omega} -
\bra{\partial_\alpha \Omega}\ketn{\Omega}\bra{\Omega}\ketn{\partial_\alpha \Omega}
\\ &= \bra{\partial_\alpha \Omega}\ketn{\Omega^t}\bra{\Omega^t}\ketn{\partial_\alpha \Omega}
\notag
=
\frac{1}{4} ({\bf e}_- \cdot \partial_\alpha {\bf n})( {\bf e}_+ \cdot \partial_\alpha {\bf n})
\\\notag  &=
\frac{1}{4}
(\partial_\alpha {\bf n}) \cdot (\partial_\alpha {\bf n}).
\end{align}
Thus the $\Upsilon$ term signifies the stiffness of the superfluid
with respect to magnetic gradients (as opposed to simply $U(1)$ phase
gradients). Also, since we identified $\Upsilon$ with the Lagrangian
density of a $CP^1$ model, we now reaffirm its equivalence
with the nonlinear sigma model \cite{rajaraman82}.

\subsection{Equations of motion for spin-half condensates}
\label{Sec:EOMhalf}

Now that we clarified how the hydrodynamic variables arise in the GP
Lagrangian density, we are ready to approach the GP equations of
motion.  In terms of the original variables, the time-dependent GPE for
a spin half condensate is
\begin{equation}
i \partial_t \psi_a = - \frac{1}{2} \nabla^2 \psi_a
+ g \rho \psi_a
\end{equation}
where we note that the interaction energy for this case is
\begin{equation}
\V = \frac{1}{2} g \rho^2.
\end{equation}
Following the substitution $\psi_a=\psi\chi_a$, with $\chi_a$ the
entries of the spin-half spinor $\ket{\chi}$, and contraction
with $\bra{\chi}$ we find
\begin{align}
\notag
i\partial_t \psi + \psi a_t = \frac{1}{2}
(-i\nabla-{\bf a})^2 \psi
+\frac{1}{2} \Upsilon\psi +g\rho\psi
\end{align}
where $a_\alpha=(a_t, {\bf a})$ is the vector potential
introduced previously.  Substituting
$\psi=fe^{i\theta}$ and multiplying both sides of the equation by
$e^{-i\theta}$ gives
\begin{eqnarray}
i\partial_t f - \partial_t \theta f + f a_t&=&
\frac{1}{2}(-\nabla^2 f -i f \nabla \cdot {\bf v} -2i {\bf v} \cdot
\nabla f + f v^2) \notag
\\
&+&
\frac{1}{2}\Upsilon f  + g \rho f
\end{eqnarray}
where ${\bf v} = \nabla \theta - {\bf a}$.
The imaginary part of this  gives
\begin{equation}
\partial_t \rho =- \nabla \cdot (\rho {\bf v})
\end{equation}
which is a mass conservation equation.
On the other hand, taking the real part gives
\begin{equation}
\partial_t \theta + \frac{1}{2} v^2 -a_t
=
\frac{1}{2} \frac{\nabla^2 f}{f} -\frac{1}{2}\Upsilon - g\rho.
\end{equation}
We take the gradient of both sides of this equation
(using the identity
$\nabla(v^2)= 2({\bf v}\cdot \nabla) 
{\bf v}+ 2{\bf v} \times (\nabla \times {\bf v})$) to get
\begin{equation}
D_t {\bf v} = \e + ({\bf v} \times \b) - \nabla \left( g \rho
+ \frac{1}{2} \Upsilon -
\frac {\nabla^2 \sqrt{\rho}}{2 \sqrt{\rho}}\right).
\end{equation}
In this we have defined the ``electric'' and ``magnetic'' fields
$\e$ and $\b$ in
the usual way from the vector potential.  That is,
$\e_{\alpha} = \f_{\alpha t}$ and $\b_{\alpha}=(\nabla \times {\bf
a})_{\alpha} =\frac{1}{2} \epsilon_{\alpha \beta \gamma} \f_{\beta
\gamma}$, with $\alpha,\,\beta$ and $\gamma$ indicating space
directions, and the $\f$ tensor defined below. Also, note that we have used the material derivative  $D_t
= \partial_t + {\bf v} \cdot \nabla$.  The ``electromagnetic force" appearing in the 
right-hand-side of the Euler equation is a new feature that is not present in single 
component condensates.  This new type of quantum pressure arises from non-uniform
spin textures in spinor condensates.

Now we move on to find the equations describing the spin dynamics.
To do this, we contract the GPE with the time reversed
spinor $\bra{\chi_t}$.  This gives
\begin{align}
i\bra{\chi_t}\ketn{D_t \chi} &= -\frac{1}{2} (
2i {\bf a} \cdot \bra{\chi_t} \ketn{\nabla \chi} +
2\frac{\nabla f}{f} \cdot \bra{\chi_t}\ketn{\nabla\chi}
\notag
\\ &
\qquad \qquad+
\bra{\chi_t}\ketn{\nabla^2 \chi})).
\end{align}
Using the spin identities developed in Sec.~\ref{Sec:spinrep}
the following relations can be derived
$\bra{\chi_t}\ketn {\nabla^2 \chi}=
\frac{1}{2} \partial_\alpha ({\bf e}_+ \cdot \partial_\alpha {\bf n})$
and $\bra{\partial_\alpha \chi}\ketn{\chi} \bra{\chi_t}
\ketn{\partial_\alpha \chi}= -\frac{1}{4}
\partial_\alpha {\bf e}_+ \cdot \partial_\alpha {\bf n}$.
In terms of vectors, the above equation is then
\begin{equation}
\frac{i}{2} {\bf e}_+ \cdot D_t {\bf n} =
-\frac{1}{2} \left( \frac{1}{2} {\bf e}_+ \cdot \nabla^2 {\bf n}
+ \frac{\partial_\alpha f}{f} {\bf e}_+ \cdot \partial_\alpha {\bf n}
\right)
\end{equation}
which can be rewritten as
\begin{equation}
\rho D_t {\bf n} = \frac{1}{2} \left( {\bf n} \times
\partial_\alpha ( \rho \partial_\alpha {\bf n})\right)
\end{equation}
which is a Landau-Lifshitz equation. 

Thus, collecting everything, we can write down a complete set of equations
describing the dynamics of a spin half condensate:
\begin{align}
-\partial_t \rho &= \nabla \cdot (\rho {\bf v}) \nonumber
\\ \nonumber
\nabla \times {\bf v} &= -{\b}
\\ \nonumber
D_t {\bf v} &= \e + ({\bf v} \times \b) - \nabla \left( g \rho
+ \frac{1}{8} (\partial_\alpha {\bf n})^2 -
\frac {\nabla^2 \sqrt{\rho}}{2 \sqrt{\rho}}\right)
\\\nonumber
\rho D_t {\bf n} &= \frac{1}{2} \left( {\bf n} \times
\partial_\alpha ( \rho \partial_\alpha {\bf n})\right)
\end{align}
where $\e$ and $\b$ are related to the spin direction through the
field tensor
\be
\f_{\alpha \beta}=
\left(\ba{cccc}
0 & -\e_x & -\e_y & -\e_z \\
\e_x & 0 & \b_z & -\b_y \\
\e_y & -\b_z & 0 & \b_x \\
\e_z & \b_y & -\b_x & 0 \\
\ea\right)
=
-\frac{1}{2} {\bf n} \cdot (  \partial_\alpha{\bf n}
\times \partial_\beta{\bf n}).
\ee
It is interesting to compare these results with those obtained
previously in the incompressible regime \cite{lamacraft08},
Eqns.~(\ref{Eq:lam}).  We find that lifting the incompressibility constraint
leads to an Euler equation with effective electric and magnetic
fields given by the Mermin-Ho relation.  
In addition, the superfluid density now enters the 
Landau-Lifshitz equation.

\subsection{Application: skyrmion texture}
\label{Sec:skyr}

As an example of the efficiency of the above hydrodynamic equations of
motion, let us consider a specific calculation: skyrmion textures in ferromagnetic condensates.
For a standard $U(1)$ vortex,
the superfluid velocity  close to the vortex core diverges as $1/r$.
For a scalar condensate, this causes the superfluid
density to be depleted in a small region
of order of the coherence length around the core.  This can be energetically
costly if the condensate is near the incompressible regime.   On the other
hand, this situation can be circumvented for a spinor condensate.
Consider for example, a two component (spin-half) condensate
$(\psi_\uparrow, \psi_\downarrow)$, and take the $\downarrow$ component to
have a $U(1)$ vortex.  Then around the vortex core, the density of
$\psi_\downarrow$ can be transferred to the vortex-free $\psi_\uparrow$
keeping the total density across the vortex core finite.  This
is known as the skyrmion configuration which has been argued to be
the relevant topological defect for ferromagnetic condensates
\cite{ho98,khawaja01,mueller04}.

\begin{figure}
\includegraphics[width=3.5in]{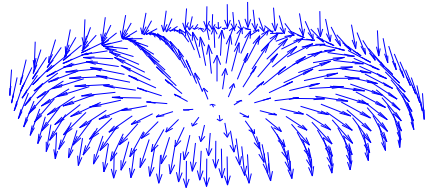}
\caption{A skyrmion configuration corresponding to Eq.~(\ref{Eq:nskyr}).}
\label{Fig:skyr}
\end{figure}

Let us now derive the
analytic time-independent solution of the equations of motion
in the incompressible regime
having the skyrmion texture shown in Fig.~\ref{Fig:skyr}.
To this end, we take the incompressible limit \cite{lamacraft08} 
of the equations of motion
for the spin-half condensate
obtained  in Sec.~\ref{Sec:EOMhalf}.
Neglecting $z$-dependence, these are:
\begin{align}
\label{Eq:in1}
\nabla \cdot {\bf v} &=0
\\
\label{Eq:in2}
\partial_x v_y - \partial_y v_x &= \frac{1}{2} {\bf n} \cdot
(\partial_x {\bf n} \times \partial_y {\bf n})
\\
\label{Eq:in3}
D_t {\bf n} &= \frac{1}{2} ({\bf n} \times \nabla^2 {\bf n}).
\end{align}
With small modifications, these equations can also be shown
to describe the dynamics of condensates confined to the ferromagnetic
phase of arbitrary
spin in the incompressible regime.  Our aim is to find stationary
solutions of these equations having a skyrmion texture given
by
\cite{ho98, mueller04}
\begin{equation}
\label{Eq:nskyr}
{\bf n} = (\sin(\beta)\cos(\varphi),
\sin(\beta) \sin(\varphi),\cos(\beta))
\end{equation}
where $\varphi$ is the
azimuthal angle and $\beta$ is a function of $r$ which is subject
to the boundary conditions $\beta(r=0)=0$ and $\beta(r=R)=\pi$ where $R$ is a distance far from the skyrmion center.  Such a spin
configuration is shown in Fig.~\ref{Fig:skyr}.

Given the form ${\bf n}$ in Eq.~(\ref{Eq:nskyr}),
Eqns.~(\ref{Eq:in1},\ref{Eq:in2})  can be solved to obtain the
velocity profile.  One finds
\begin{equation}
\label{Eq:vskyr}
{\bf v} = \frac{\sin^2(\beta/2)}{r} \hat{\varphi}.
\end{equation}
Note that the boundary condition $\beta(0)=0$ suppresses the
velocity at the origin which diverges as $1/r$ for
the standard $U(1)$ vortex.
With the assumption of a static configuration, Eq.~(\ref{Eq:in3})
reduces to
\begin{equation}
\label{Eq:statLL}
{\bf v} \cdot \nabla {\bf n} = \frac{1}{2} ({\bf n} \times \nabla^2 {\bf n}).
\end{equation}
With the expression for ${\bf v}$ in Eq.~(\ref{Eq:vskyr}),
Eq.~(\ref{Eq:statLL})
leads to the following second order differential equation for $\beta$
\begin{equation}
r \left(r \frac{d^2 \beta}{dr} + \frac{d\beta}{dr} \right)=\sin(\beta).
\end{equation}
With the boundary conditions, the solution of this differential equation is
\begin{equation}
\beta(r)= 4 \tan^{-1}\left(r/R\right).
\end{equation}
This expression, along with the velocity in Eq.~(\ref{Eq:vskyr}) and
the spin direction in Eq.~(\ref{Eq:nskyr}) constitute an
analytic stationary solution to the equations of
motion for the skyrmion configuration.

\section{Hydrodynamics of spin-one condensates}
\label{Sec:Fone}

\subsection{Geometrical representation of spin one hydrodynamic quantities}

Now we move on to considering the more complicated
case of the spin-one condensate.  The spin-one 
spinor can be broken down into
its two spin-half components and be written as
$\ket{\chi}= \ket{\bo}/
\sqrt{\bra{\bo}\ketn{\bo}}$ where
$\ket{\bo} = \ket{\Omega_1 \Omega_2}$, where we again make use of the
large-spin notation defined in Eq.~(\ref{Eq:spinor}).  
The normalization
factor for this case is found to be
\begin{align}
\bra{\bo}\ketn{\bo} &= \bra{\Omega_1}\ketn{\Omega_1}
\bra{\Omega_2}\ketn{\Omega_2}+\bra{\Omega_1}\ketn{\Omega_2}
\bra{\Omega_2}\ketn{\Omega_1} \notag
\\
&=
\frac{3}{2} + \frac{1}{2} {\bf n}_1 \cdot {\bf n}_2.
\end{align}

It is also instructive to calculate the spin operator's expectation
value.  To start we can expand into products of spin-half expectation
values
\begin{align}
\bra{\bo} {\bf F} \ket{\bo} &= \bra{\Omega_1}{\bf F} \ket{\Omega_1}
+ \bra{\Omega_2}{\bf F} \ket{\Omega_2} \\ \notag &+
\bra{\Omega_1}\ketn{\Omega_2} \bra{\Omega_2}{\bf F} \ket{\Omega_1}
+
\bra{\Omega_2}\ketn{\Omega_1} \bra{\Omega_1}{\bf F} \ket{\Omega_2}.
\end{align}
Then using the identity in Eq.~(\ref{id1}), and the
fact that in the factored expression, ${\bf F}$ is only acting 
on spin-half states, we obtain
$
\bra{\bo} {\bf F} \ket{\bo}={\bf n}_1 + {\bf n}_2.
$
Dividing this by the
normalization, we get the spin-one expectation value of the magnetization:
\begin{equation}
\label{Eq:ms1}
{\bf m} = \bra{\chi} {\bf F} \ket{\chi} = 2 \frac{{\bf n}_1 + {\bf
n}_2}{3+{\bf n}_1 \cdot {\bf n}_2}.
\end{equation}

By similar techniques, the vector potential for 
the spin-one case, with some work,
can be written as
\begin{align}
a_\alpha= i\bra{\chi}\ketn{\partial \chi} &=
\frac{1}{2} {\bf e}_{1y}  \cdot \partial_\alpha{\bf e}_{1x}+
\frac{1}{2} {\bf e}_{2y}  \cdot \partial_\alpha{\bf e}_{2x}
\\ \nonumber
&+
\frac{1}{2} \frac{({\bf n}_2 \times {\bf n}_1)\cdot \partial_\alpha {\bf n}_1
+ ({\bf n}_1 \times {\bf n}_2)\cdot \partial_\alpha {\bf n}_2}
{3 + {\bf n}_1 \cdot {\bf n}_2}.
\end{align}
One sees that the first two terms in this expression are the vector
potentials from the individual spin-half components while the final
term, which is gauge invariant, describes their coupling.  This expression
was previously obtained in Refs. \cite{bouchiat88} and 
\cite{hannay98}, where a geometrical relation for the Berry phase
of a spin-one spinor was given.
The field tensor corresponding to this vector potential
can also be similarly computed.  The most simplified form we find is
\begin{align}
\label{Eq:MHs1}
\f_{\alpha \beta}&=\frac{-2}{(3+{\bf n}_1 \cdot {\bf n}_2)^2} \times
\\&(2{\bf
n}_1 \cdot (\partial_\alpha{\bf n}_1 \times \partial_\beta {\bf
n}_1)
+2{\bf n}_2 \cdot(\partial_\alpha{\bf n}_2 \times
\partial_\beta {\bf n}_2) \notag
\\ \nonumber
&+
({\bf n}_1+{\bf n}_2) \cdot(\partial_\alpha{\bf n}_1 \times
\partial_\beta {\bf n}_2+\partial_\alpha{\bf n}_2 \times
\partial_\beta {\bf n}_1) ).
\end{align}
This is a generalization of the Mermin-Ho relation to the spin-one case.
To our knowledge such an expression has not been previously derived.
While its geometrical interpretation is not as immediate as the 
spin-half case (which is the Pontryagin density), this expression might
be of use in computing topological invariant quantities for spin-one
fields.  This formula has a simplified form when locally restricted
to mean-field ground states.  For instance for the ferromagnetic sate
(${\bf n} \equiv {\bf n}_1 = {\bf n}_2$)
the above expression reduces to
\begin{equation}
\f_{\alpha \beta}=-{\bf n}\cdot (\partial_\alpha{\bf n} 
\times \partial_\beta {\bf n}).
\end{equation}
It is also useful to note that for the nematic state 
(${\bf n} \equiv {\bf n}_1 = -{\bf n}_2$) the field tensor identically
vanishes, $\f_{\alpha \beta}=0$.

Finally, the gauge invariant quantity $\Upsilon$
can be worked out to be
\begin{align}
\label{Eq:Upsilons1}
\Upsilon &=\frac{2}{(3+{\bf n}_1 \cdot {\bf n}_2)^2}(
\partial_\alpha {\bf n}_1 \cdot \partial_\alpha {\bf n_1}+
\partial_\alpha {\bf n}_2 \cdot \partial_\alpha {\bf n_2}+
\partial_\alpha {\bf n}_1 \cdot \partial_\alpha {\bf n_2}&
\notag
\\
&+ {\bf n}_1 \cdot {\bf n_2} \; \partial_\alpha {\bf n}_1 \cdot \partial_\alpha
{\bf n_2}- {\bf n}_1 \cdot \partial_\alpha {\bf n}_2 \;{\bf n}_2 \cdot
\partial_\alpha {\bf n_1}).
\end{align}
This is an explicit representation of the CP$^2$ model which can
be viewed as a generalization of the nonlinear sigma model.
Here, too, it is instructive to consider what this expression reduces
to when locally restricting to mean-field ground states.  For the
ferromagnetic state, one finds
\begin{equation}
\Upsilon = \frac{1}{2} \partial_\alpha {\bf n} \cdot \partial_\alpha {\bf n}. 
\end{equation}
On the other hand, for the nematic state $\Upsilon$ reduces to
\begin{equation}
\Upsilon = \frac{1}{4} \partial_\alpha {\bf n} \cdot \partial_\alpha {\bf n}. 
\end{equation}

\subsection{Spin-one condensate equations of motion}

We now  proceed to do a similar analysis for the spin one
problem.  For this we note that the spin one GP energy functional has the
form
\begin{equation}
\V=\frac{1}{2} g \rho^2 + \frac{1}{2} c_2 \rho^2 {\bf m}^2
\end{equation}
where ${\bf m}$ is the expectation value of the spin-one operator.
The first two
hydrodynamic equations -- the mass
continuity equation and the modified Euler
equation -- are obtained, as before,  by contracting the Gross Pitaevskii
equation with $\bra{\chi}$.
The analysis proceeds along similar lines as the spin half case.
However, for this case
we need the generalization
of the Mermin-Ho relation for spin one given in Eq.~(\ref{Eq:MHs1}) to give
the field tensor and
thus the effective electric and magnetic fields, in addition to the spin one
expressions for $\Upsilon$ Eq.~(\ref{Eq:Upsilons1}) and the magnetization
${\bf m}$, in Eq.~(\ref{Eq:ms1}).
With these quantities, the first two
equations of motion are
\begin{equation}
\partial_t \rho = -\nabla \cdot (\rho {\bf v})
\end{equation}
and
\begin{align}
D_t {\bf v} = \e + ({\bf v} \times \b) - \nabla \left( g
\rho + c_2 \rho m^2+ \frac{1}{2} \Upsilon 
-  \frac{\nabla^2 \sqrt{\rho}}{2 \sqrt{\rho}}\right).
\end{align}

Next, let us discuss the spin dynamical equations. These are
obtained by contracting the GPE with $\bra{\Omega_1^t\Omega_1^t}$
and $\bra{\Omega_2^t\Omega_2^t}$.  
As before, this causes several terms to vanish since these spinors
are orthogonal to $\ket{\chi}$.  Contracting with $\bra{\Omega_1^t\Omega_1^t}$
gives the following equation which gives the time derivative of the first
node
\begin{equation}
i {\bf e}_{1+} \cdot D_t {\bf n}_1 = 
 - \frac{1}{2} \Gamma_\alpha^{12}  {\bf e}_{1+} 
\cdot \partial_\alpha {\bf n}_1 
-\frac{1}{2}  {\bf e}_{1+} \cdot \nabla^2 {\bf n}_1 
+ c_2 \rho {\bf e}_{1+} \cdot {\bf m}.
\end{equation}
A similar equation for the time derivative of ${\bf n}_2$ is obtained
by contracting with $\bra{\Omega_2^t\Omega_2^t}$.  In the above,
we have collected the following terms into the 
$\Gamma^{ij}_\alpha$ parameter 
\begin{align}
\Gamma^{ij}_\alpha &= 2 \frac{\partial_a \sqrt{\rho}}{\sqrt{\rho}}-
 \frac{\partial_\alpha({\bf n}_i \cdot {\bf n}_j)}{3+{\bf n}_i \cdot {\bf n}_j}
\\\qquad
&-
\frac{{\bf n}_i \cdot \partial_\alpha {\bf n}_j - i {\bf n}_i \cdot ({\bf n}_j \times \partial_\alpha {\bf n_j})}{1-{\bf n}_i \cdot {\bf n}_j}
\notag
\\ \qquad
&+  i \frac{({\bf n}_j \times {\bf n}_i)\cdot \partial_\alpha {\bf n}_i
+ ({\bf n}_i \times {\bf n}_j)\cdot \partial_\alpha
{\bf n}_j}{3+{\bf n}_i \cdot {\bf n}_j}.
\end{align}
Finally, separating the real and imaginary parts as
$\Gamma^{ij}_\alpha = (\Gamma^{ij}_\alpha)'+i(\Gamma^{ij}_\alpha)''$,
we obtain the Landau-Lifshitz equations 
\begin{equation}
(D_t +\frac{1}{2} (\Gamma^{12}_\alpha)'' \partial_\alpha){\bf n}_1
= \frac{1}{2} {\bf n}_1 \times ((\Gamma^{12}_\alpha)' \partial_\alpha {\bf n}_1 +
\nabla^2 {\bf n}_1)- c_2 \rho {\bf n}_1 \times {\bf m}
\end{equation}
\begin{equation}
(D_t +\frac{1}{2} (\Gamma^{21}_\alpha)'' \partial_\alpha){\bf n}_2
= \frac{1}{2} {\bf n}_2 \times ((\Gamma^{21}_\alpha)' \partial_\alpha
{\bf n}_2 +
\nabla^2 {\bf n}_2)- c_2 \rho {\bf n}_2 \times {\bf m}
\end{equation}
This provides a complete set of equations describing the dynamics of the
spin-one condensate.

\section{Hydrodynamics for general spin-$F$ condensates.}
\label{Sec:genF}

Now that we have considered the hydrodynamic equations for 
spin-half and spin-one condensates in detail, in the following
we will consider the general case.  
The first two equations of motion, the mass continuity equation and
the Euler equation are found, as before,  to be
\begin{equation}
\partial_t \rho = -\nabla \cdot (\rho {\bf v})
\end{equation}
and
\begin{align}
D_t {\bf v} = \e + ({\bf v} \times \b) - \nabla \left(
\frac{2 \V}{\rho}+ \frac{1}{2} \Upsilon 
 -\  \frac
{\nabla^2 \sqrt{\rho}}{2 \sqrt{\rho}}\right).
\end{align}
The effective electric and magnetic
fields again follow from the field tensor $f_{\alpha \beta}$
constructed from $a_{\alpha}=i\bra{\chi}\ketn{\partial_\alpha
  \chi}$. For a general spin, however, such quantities are cumbersome to
express directly in terms of the spin nodes, and we will refrain from doing so.

To obtain the Landau-Lifshitz equations, we contract
the GPE with $\bra{\cotr{i}}$.  Doing this gives
\begin{align}
\notag
i\bra{\Omega_i^t}\ketn{\partial_t \Omega_i} &=
- \partial_\alpha \log\left(\frac{\psi}{\sqrt{\bra{\bo}\ketn{\bo}}}\right)
\bra{\Omega_i^t}\ketn{\partial_\alpha \Omega_i} \\ \notag&- 
-\frac{1}{2}\bra{\Omega_i^t}\ketn {\nabla^2 \Omega_i}
-
\bra{\Omega_i^t}\ketn{\partial_\alpha \Omega_i}
\sum_{j\ne i} \frac{\bra{\Omega_i^t}\ketn{\partial_\alpha \Omega_j}}{\bra{\Omega_i^t}\ketn{\Omega_j}}
\\
\label{Eq:LLgen}
&+ \frac{\rho}{\lambda_i^* \bra{\bo} \ketn{\bo}}
\brasub{\cotr{i}}{1} \brasub{\bo}{2} {\cal \V} \ketsub{\bo}{2}\ketsub{\bo}{1}.
\end{align}
In this expression, we have used the notation for interaction energy introduced
in Eq.~(\ref{Eq:spinintalt}).  In addition we have introduced the
quantities $\lambda_i$,
\begin{equation}
\lambda_i = (2F)!\prod_{j\ne i} \bra{\Omega_j} \ketn{\Omega_i^t}.
\end{equation}

Instead of writing Eq.~\ref{Eq:LLgen} in terms of vectors as in the previous
sections, we will stop at this point.  This equation provides a natural
starting point in the analysis of the linearized equations of motion
which will be developed in the companion paper Ref.~[\onlinecite{barnett09}].

\section{Conclusions}

One of the most striking and surprising features of spinor
condensates is the hidden symmetry of their mean field ground states. In
this work, we have strived to bring this symmetry to the forefront,
and describe hydrodynamics of these condensates using the objects that
exhibit the hidden symmetry, namely, the spin-nodes, and reciprocal
coherent states.

In this work, we derived the hydrodynamic equations of motion for condensates of general spin, 
demonstrated their use in the computation of the skyrmion 
configuration of a ferromagnetic spin-half gas, and
generalized the Mermin-Ho relation to spin-one condensates.

A hydrodynamic description tries to capture the low-energy behavior of a
continuous medium in terms if conserved and other simple
quantities. Therefore, it comes particularly handy when we consider
spinor-condensates close to their mean-field ground state. 
In a companion paper \cite{barnett09}, we will concentrate on small 
oscillations of the spinor
condensate in the vicinity of the mean field-ground state. As we will show, it is there
that the hidden point-group symmetry becomes most apparent and
accessible. Using the spin-node formalism, and the parametrization 
of the spin-nodes in terms of a stereographic
projection, we will reduce the problem
of finding the $2F$ spin-wave eigenmodes to a simple question of
decomposing a representation of the appropriate point symmetry group to its
irreducible representations. We will also provide a simple recipe that 
allows the direct extraction of the
condensate's spin-wave eigenmodes using the derivatives of the 
spherical harmonics,
coupled with the knowledge of atomic orbital degeneracy lifting under
a crystal field.

\acknowledgments

It is a pleasure to acknowledge useful conversations with E. Demler, 
T.-L. Ho, I. Klich, A. Lamacraft, 
and especially A. Turner. We would like to acknowledge the hospitality
of the KITP, 
supported by NSF
PHY05-51164.
We are also grateful for support from 
the Sherman Fairchild Foundation (RB); the Packard and Sloan
Foundations, and the Institute for Quantum 
Information under NSF grants PHY-0456720 and
PHY-0803371 (GR); and CIFAR, NSERC, and CRC (DP).

\appendix

\section{Notation}
\label{A1}
In this Appendix, for convenience, we collect in one place the
notation used for the representation of the spinors.
Normalized spinors of arbitrary spin are denoted by
$\ket{\chi}$.  Non-normalized symmetric combinations of
spin-half states are denoted as (with bold fonts)
\begin{equation}
\ket{\bo} = \ket{\Omega_1 \Omega_2 \ldots \Omega_{2F}}.
\end{equation}
This is then related to $\ket{\chi}$ by
\begin{equation}
\ket{\chi} = \frac{\ket{\bo}}{\sqrt{\bra{\bo}\ketn{\bo}}}.
\end{equation}
Coherent spin states occur when all of the spin-half constituent
spins point in the same direction.  We denote these by
\begin{equation}
\ket{\co{i}}=\ket{\Omega_i \Omega_i \ldots \Omega_i}.
\end{equation}
We next define the spin state corresponding
to $\ket{\bo}$ with its $ith$ component time-reversed.  We
denote these by
\begin{equation}
\ket{T_i {\bo}}=\ket{\Omega_1 \Omega_2 \ldots
\Omega_{i}^t \ldots \Omega_{2F}}.
\end{equation}
Finally, we define the projection operator $\Pn$ to be
\begin{equation}
\Pn = 1- \ket{\chi} \bra{\chi}.
\end{equation}


\end{document}